\documentclass[10pt,preprint]{aastex}

\begin{document}

\newcommand{\kms}{\,km\,s$^{-1}$}

\title{FUSE Observations of \ion{O}{6} Absorption in the Galactic Halo}

\author{B.D.~Savage\altaffilmark{1}, K.R.~Sembach\altaffilmark{2},
        E.B.~Jenkins\altaffilmark{3},
        J.M.~Shull\altaffilmark{4},
        D.G.~York\altaffilmark{5},
        G. Sonneborn\altaffilmark{6},
        H.W.~Moos\altaffilmark{2},
        S.D. Friedman\altaffilmark{2},
        J.C. Green\altaffilmark{4},
        W.R. Oegerle\altaffilmark{2},
        W.P. Blair\altaffilmark{2},
        J.W. Kruk\altaffilmark{2},
        E.M. Murphy\altaffilmark{2}}
\altaffiltext{1}{Department of Astronomy, University of Wisconsin, Madison,
        WI  53706}
\altaffiltext{2}{Department of Physics \& Astronomy, The Johns
        Hopkins University, Baltimore, MD  21218}
\altaffiltext{3}{Princeton University observatory, Princeton, NJ 08544}
\altaffiltext{4}{CASA and JILA, Department of Astrophysical \& Planetary Sciences,
        University of Colorado, Boulder, CO 80309}
\altaffiltext{5}{Astronomy \& Astrophysics Dept., University of Chicago,
        Chicago 60637}
\altaffiltext{6}{NASA/GSFC, Greenbelt, MD 20771}

\begin{abstract}
Far-ultraviolet spectra of 11 AGNs observed by FUSE are analyzed to
obtain measures of
\ion{O}{6} $\lambda$1031.93 absorption  occurring over very long  paths through
Milky Way halo gas. Strong \ion{O}{6} absorption is detected along 10 of 11
sight lines.  Values of $\log$[N(\ion{O}{6}) $\sin|$b$|$] range from
13.80 to 14.64 with a median
value of 14.21. The observations reveal the existence of a widespread but
irregular distribution of \ion{O}{6} in the Milky Way halo.  Combined
with estimates of
the \ion{O}{6} mid-plane density, $n_0$ = $2\times10^{-8}$ cm$^{-3}$ ,
from the
{\it Copernicus} satellite, the FUSE observations imply an \ion{O}{6}
exponential scale
height of 2.7$\pm$0.4 kpc.  We find that N(\ion{C}{4})/N(\ion{O}{6}) ranges
from $\sim$0.15 in the disk to $\sim$0.6 along four extragalactic sight lines.
The changing ionization state of the gas
from the disk to the halo is consistent with a systematic decrease in the
scale heights of \ion{Si}{4}, \ion{C}{4}, \ion{N}{5},
to \ion{O}{6} from $\sim$5.1 to $\sim$2.7 kpc.  While conductive heating models
can account for the highly ionized atoms
at low $|$z$|$, a combination of models (and processes) appears to be
required to explain the highly ionized
atoms found in the halo.  The greater scale heights of \ion{Si}{4} and \ion{C}{4}
compared to \ion{O}{6} suggests that
some of the \ion{Si}{4} and \ion{C}{4} in the halo is produced
in turbulent mixing layers or by
photoionization by hot halo stars or the extragalactic background.

\end{abstract}


\section{Introduction}
In the fundamental paper ``On a Possible Interstellar Galactic Corona,''
Spitzer (1956) discussed the physical basis for the possible existence
of a
hot interstellar gas phase extending away from the Galactic plane into the
halo.  He also noted the diagnostic potential of the resonance doublet
absorption lines of the Li-like ions of \ion{O}{6}, \ion{N}{5}, and
\ion{C}{4} for studying the coronal Galactic halo gas.  Under conditions of
equilbrium collisional ionization, these three ions peak in abundance at
$(3, 2, 1)\times10^5$ K, respectively (Sutherland \&  Dopita 1993).

Interstellar \ion{O}{6} observations were obtained by
Jenkins (1978a, b) and York (1977) using the high resolution far-UV
spectrometer aboard
the {\it Copernicus} satellite.  Those observations provided fundamental
information about the hot interstellar gas in the disk of the Milky Way
but were limited to stars brighter than V $\sim$ 7.0.  The {\it International
Ultraviolet Explorer} (IUE) and the {\it Hubble Space Telescope}
(HST) allowed astronomers to study absorption by \ion{N}{5}, \ion{C}{4},
and \ion{Si}{4}
in the Galactic halo
(Savage \& de~Boer 1979; Savage \& Massa 1987; Sembach \&
Savage 1992; Savage, Sembach, \& Lu 1997), but \ion{O}{6} was unobservable
because windowless detectors and specially coated optics
are required to measure the \ion{O}{6}
$\lambda\lambda$1031.93, 1037.62 doublet.  \ion{O}{6} has a special significance
among the high ionization species because
it is a sensitive indicator of collisionally ionized gas and is the
least likely
to be produced by photoionization from starlight given that 113.9 eV are
required for the converison of \ion{O}{5} to \ion{O}{6}.

Except for brief observing programs by the {\it Hopkins Ultraviolet
Telescope} (HUT; Davidsen 1993), and the spectrographs in the {\it Orbiting
and Retrievable Far and Extreme Ultraviolet Spectrometers} (ORFEUS;
Hurwitz \& Bowyer 1996; Hurwitz et al. 1998; Widmann et al.
1999; Sembach, Savage, \& Hurwitz 1999), we have had a long hiatus in
observing \ion{O}{6}.  This has ended with the commissioning of the
{\it Far-Ultraviolet
Spectroscopic Explorer} (FUSE), a facility that has been
specially designed to cover wavelengths from 905 to 1187~\AA\ with an efficient
2-dimensional detector.  The
spectroscopic capabilities of FUSE and its in-flight performance are
discussed by Moos et al.\ (2000) and Sahnow et al.\ (2000).  This paper reports
on initial FUSE results for \ion{O}{6} absorption in the Galactic halo.

\section{FUSE Observations and Data Processing}
The spectral integrations with FUSE were obtained in the time-tagged photon
address mode between September and December 1999.   The observations were
obtained with the objects centered in the large
30\arcsec$\times$30\arcsec\ aperture of the LiF1 channel
and generally extended over spacecraft
night and day.  Most of the effective area in the \ion{O}{6}
$\lambda\lambda$1031.93,
1037.62 wavelength region is provided by the spectra obtained with the LiF
channels.  We have restricted our analysis to the observations in the
LiF1 channel.

To produce the final composite spectra in the region of \ion{O}{6}, we
followed the
basic data handling procedures discussed by Sembach et al.\ (2000).  The
spectra have a
velocity resolution of approximately 25\kms\ (FWHM). The zero point of
the wavelength scale in
the vicinity of the \ion{O}{6} lines was determined with reference to
nearby H$_2$ lines or
\ion{Si}{2} and \ion{O}{1} lines.  Sample FUSE spectra  extending  from
1020  to  1045~\AA\  for  Ton~S210  and  PG~0804+761  are  shown  in 
Figure  1.

\section{Interstellar Analysis}
The weaker \ion{O}{6}  $\lambda1037.62$  line  is  near strong absorption
by  \ion{C}{2}$^*$ $\lambda1037.02$  and  the  H$_2$  (5--0) R(1) and P(1)
lines  at 1037.15 and  1038.16~\AA.  The stronger  \ion{O}{6}  $\lambda1031.93$
line  is usually  relatively  free  of  blending  with  other  species
since  the nearby  H$_2$  (6--0) R(3) and R(4) lines at 1031.19
and  1032.36~\AA\ are  often  weak  and  well-separated  in
velocity  from \ion{O}{6}  ($\Delta v$  = --214  and  +125 \kms, respectively).
Blending from the HD (6--0) R(0) and R(1) lines at 1031.93 and 1032.36~\AA\
is not a problem because the amount of molecular gas along these sight lines
is small (Shull et al. 2000).  For most paths through the Galactic
halo, the contamination of the \ion{O}{6}  $\lambda1037.62$  line  is severe.
Therefore, we concentrate our attention on the \ion{O}{6}
$\lambda1031.93$  line
in this Letter and use the $\lambda1037.62$ line only to evaluate
possible saturation
in the $\lambda1031.93$ absorption.

The \ion{O}{6} $\lambda1031.93$ absorption is sufficiently
broad that it is nearly fully resolved by FUSE.  Therefore,
we converted the observed absorption line profiles into measures of \ion{O}{6}
apparent column density per unit velocity, N$_a$($v$), through the
relation N$_a(v)$ [{\rm ions\ cm}$^{-2}$ ({\rm km\,s}$^{-1}$)$^{-1}$]
= $m_ec/\pi e^2$ $\tau_{a}(v) (f\lambda)^{-1}$=3.768$\times$10$^{14}$
$\tau_a(v) (f\lambda)^{-1}$, where $f$=0.133 is the oscillator
strength  of  the  $\lambda1031.93$ line (Morton 1991), $\lambda$ is the
wavelength  in \AA,  and  $\tau_a$($v$) is  the  apparent  absorption  optical
depth  (see  Savage \&  Sembach  1991).  The  continuum  levels are
well-defined by the smooth  flux  distributions  provided  by  the AGNs.
 For
cases where  the  \ion{O}{6}  $\lambda$1037.62  line  could   be 
measured,  we  find  the
same  values  of  $\log$N$_a$($v$) near  maximum  absorption  as  those
obtained  for  the  stronger  $\lambda$1031.93  line.     Therefore,
there  is  no evidence  for  unresolved  saturated  structure  in  the  \ion{O}{6}
absorption.

Values  of the integrated  apparent  \ion{O}{6}  column  density  are  given
in  Table~1  along  with  errors  including  statistical  and  continuum
placement  uncertainties (Sembach  \& Savage 1992). Column  densities
and equivalent widths
were integrated over  the  velocity  range spanned by $v_-$   to  $v_+$
as listed  in
Table~1.

\section{\ion{O}{6} Profiles}
The \ion{O}{6} absorption profiles illustrated in Figure 1 and in
Sembach et al.\ (2000) are often complex
and trace a wide range of phenomena in or
near the Milky Way.  Those portions
of the profiles between approximately --100 and  +100 \kms\
are likely tracing gas in the thick \ion{O}{6} halo of the Milky Way,
which is
the primary subject of this paper.   However, an inspection of the
\ion{O}{6} profiles reveals high velocity (100 $< |v| <$ 300 \kms\ )
\ion{O}{6} absorption toward 7 of the 12 objects: Mrk~876, Mrk~509,
PKS~2155-304, H~1821+643,
NGC~7469, Ton~S180, and Ton~S210. Sembach et al.\ (2000) discuss the
properties and
possible origins of this high velocity \ion{O}{6} absorption. We comment
on one case here.
The \ion{O}{6} absorption toward H~1821+653 reveals local gas at $v$ =
0, gas above the
Perseus spiral arm at $v$ = --70 \kms, and gas in the outer warped
region of
the outer Milky Way at $v$ = --120 \kms\ (Oegerle et al. 2000).
In calculating the column densities listed in Table 1, the velocity
limits were set to exclude high velocity \ion{O}{6} except
for H~1821+643, where the value of N(\ion{O}{6}) includes the outer
Galaxy gas
absorption extending to $v$ = --160 \kms.  This high velocity \ion{O}{6} absorption
is likely produced by differential Galactic rotation, which causes the absorption
to extend to large negative velocities.

\section{The Distribution of \ion{O}{6} in the Halo}

Total column densities of \ion{O}{6} through the Galactic halo coupled with
estimates of the mid-plane space density of \ion{O}{6}, $n_0$, can be
used to obtain
information about the stratification of \ion{O}{6} away from the plane
of the Galaxy.
We assume an exponential gas stratification with  $n(|z|)$ = $n_0 \exp(-|z|/h)$,
where $h$ is the \ion{O}{6} scale height.  It then follows that the \ion{O}{6}
column density perpendicular to the plane for an object with latitude
$b$ is
given by  N(\ion{O}{6})$\sin|b|$  = $n_0 h [1 - exp(-|z|/h)$] , where
N(\ion{O}{6}) is
the line-of-sight column density to an object at a distance $|z|$ away
from the plane.   For extragalactic objects, where $|z|\gg h$,
N(\ion{O}{6})$\sin|b|$ =
$n_0 h$.

Values of $\log$[N(\ion{O}{6})$\sin|b|$] toward 11 extragalactic objects
observed by FUSE are
listed in Table~1 along with the
ORFEUS value for 3C~273 from Hurwitz et al.\ (1998).  The large value
(14.80) for the 3C~273 sight line is likely a
consequence passing through Radio Loops I and IV and the
North Polar Spur. Such structures are local examples of the regions that likely
feed hot gas into the halo.  Since the 3C~273
sight line is a special case, we do not include it in our synthesis of
general conclusions below.  Without 3C~273, the median $\log$[N(\ion{O}{6})$\sin|b|$]
in our sample is 14.21 with a spread of 0.4 dex.  The irregular nature
of the distribution
is highlighted by the low value of 13.80 found toward VII~Zw118 which
lies $\sim14$ degrees
from PG~0804+761 and has $\log$[N(\ion{O}{6})$\sin|b|$] = 14.21.  These
irregularities must
be considered when estimating the \ion{O}{6} scale height.

From the {\it Copernicus} observations of \ion{O}{6} absorption toward
hot stars, Jenkins
(1978b) estimated a mid-plane density  $n_0$ = $2.8\times10^{-8}$ cm$^{-3}$.
However, that estimate assumed a small scale height (0.3 kpc)
for the \ion{O}{6} absorbing layer, based on the limited data available
at the
time.  We now find that the scale height must be about 10 times larger.
Accordingly, we have  obtained  a  new  estimate  of
$n_0$  from  the  {\it Copernicus}  \ion{O}{6}  survey  that  is 
appropriate  for  a
large  scale  height  \ion{O}{6}  absorbing  layer  with  h $>$ 2  kpc. 
   The
result  is  $n_0$ = $\Sigma$\,N(\ion{O}{6}) / $\Sigma$\,$r_0$ = $2.0\times10^{-8}$
cm$^{-3}$,  where the  reduced  distance  $r_0$ = $h$[$1 - \exp(-|z|/h)$]$\csc|b|$,
compensates for the small reduction in density away from the Galactic
plane.  This value for
$n_0$ was obtained when all of the upper limits for the {\it Copernicus} \ion{O}{6}
column densities were included at their stated values.  If zero is
substituted for these cases, a value of $n_0$ that is only 5\% lower is
obtained.  This value of $n_0$ has not been adjusted to allow for the fact
what we live in the Local Bubble.  Shelton \& Cox (1994) have reanalyzed
the {\it Copernicus} \ion{O}{6} measurements
and have estimated that the mid-plane \ion{O}{6} density beyond the
Local Bubble
is $n_0$ = $(1.3-1.5)\times10^{-8}$ cm$^{-3}$ for an \ion{O}{6}
absorbing layer
with h $\sim$ 3 kpc.

For an \ion{O}{6} mid-plane density of $2.0\times10^{-8}$ cm$^{-3}$, we obtain
the scale heights listed in Table~1 from the simple relation
h =  N(\ion{O}{6})$\sin|b|/n_0$.  Ignoring 3C~273, the values range
from 1.0 to 7.0 kpc.  We find median and average values of 2.6 and 2.9 kpc,
respectively.  In deriving this average and in the subsequent calculations
we treat the upper limit for PG~0052+251 as a detection.  If we adopt
the Shelton \& Cox (1994) mid-plane density estimate of $1.4\times10^{-8}$
cm$^{-3}$ and reduce the extragalactic
column densities by $1.5\times10^{13}$ cm$^{-2}$ to remove the Local Bubble
contribution, we obtain median and average
\ion{O}{6} scale heights of 3.5 and 4.0 kpc.  Our incomplete knowledge
of the
mid-plane  density introduces a 35\% systematic uncertainty in the
derivation of the \ion{O}{6} scale height.  Another source of
uncertainty involves the irregular distribution of the gas.  Edgar \& Savage
(1989) devised an analysis procedure for estimating scale heights that
accounts for the irregular distribution.  The analysis includes a
logarithmic patchiness parameter,
$\sigma_p$, that  is  added  in  quadrature to  the
 observed  logarithmic  errors  in  the  column  densities.
The  value of  $\sigma_p$ is  varied  until  the  minimized  reduced
chi  square,   $\chi^2_\nu$(min), of  the  scale  height  fit  is acceptable.
Using n$_0$  =  $2.0\times10^{-8}$ cm$^{-3}$ and  the  11  FUSE  values of
N(\ion{O}{6}) sin$|$b$|$, we  obtain $\chi^2_\nu$(min) = 1.0  for h  =  
2.7  kpc
and  $\sigma_p$ = 0.21 dex.  Adopting  this value  of  $\sigma_p$, we 
can  then estimate
the  1$\sigma$ error in  the  scale  height  by determining  the  values
 of  h  where  $\chi^2$ =
$\chi^2$(min) + 1.0.  The  final  result  is  h(\ion{O}{6}) =
2.7$\pm$0.4 kpc,
where  the listed 1$\sigma$ errors do not include the additional 35\%
systematic uncertainty
caused by the uncertain \ion{O}{6} mid-plane density and Local Bubble correction.

The \ion{O}{6} scale height of 2.7$\pm$0.4 kpc can be compared with the
values of h = 5.1$\pm$0.7, 4.4$\pm$0.6, and 3.9$\pm$1.4 kpc for
\ion{Si}{4}, \ion{C}{4},
and \ion{N}{5}, respectively, determined by Savage et al.\ (1997) from
HST and IUE
observations.  The more confined distributions of
\ion{O}{6} and \ion{N}{5} compared to \ion{C}{4} are also apparent in
plots of
N(\ion{C}{4})/N(\ion{O}{6}) and N(\ion{C}{4})/N(\ion{N}{5}) versus $|z|$
toward objects
in the disk and halo of the Galaxy.  In each case there is a clear
increase in the
ratio from the disk to the halo, suggesting that \ion{C}{4} is more extended
than \ion{O}{6} and \ion{N}{5}.  For \ion{N}{5} the ratio increases by
about a factor of
2 from low to high $|z|$  (see Fig. 6d in Savage et al. 1997).
Spitzer (1996) noted that the value of N(\ion{C}{4})/ N(\ion{O}{6})
increases from $\sim$0.15 in the disk to 0.9 for objects in the low halo with
$|z| \sim$ 1.5 kpc.   Values of N(\ion{C}{4})
have been measured using HST for Mrk~509, PKS~2155-304, 3C~273,
H~1821+643, and ESO~141-55 by Savage et al.\ (1997) and by Sembach et
al.\ (1999).
Combining these with the values of N(\ion{O}{6}) from Table 1,  we obtain
N(\ion{C}{4})/N(\ion{O}{6}) = 0.58, 0.63, 0.45, 0.63 and 1.74,
respectively, for the five extragalactic objects.   The increase in
this ratio by about a factor of 4 from the disk to the typical extragalactic
halo sight line implies a large change in the ionization state of the
highly ionized gas as a function of distance from the Galactic plane.

A  major  goal  of  the  FUSE  \ion{O}{6}
program is to map out the distribution of \ion{O}{6} in the disk and
halo of the
Galaxy.  Once we obtain FUSE \ion{O}{6} observations for a substantial
number of disk and halo stars and for additional extragalactic objects it
will be possible to improve on our intitial estimate of the extension of
\ion{O}{6} into
the Galactic halo.  These future studies will allow us to determine if a
plane parallel,
exponentially stratified, and patchy layer is indeed the most
appropriate description of the
distribution of \ion{O}{6}.

\section{The Origin of Highly Ionized Atoms in the Galactic Halo}
Strong \ion{O}{6} absorption toward 10 of 11 extragalactic
objects observed by FUSE implies the widespread existence of hot gas in
the halo of the Milky
Way as predicted by Spitzer (1956) and also supported by the earlier
observations of \ion{C}{4} and \ion{N}{5} with IUE and HST.  The
decreasing scale
heights in the sequence \ion{Si}{4}, \ion{C}{4}, \ion{N}{5}, to \ion{O}{6}
provides information about the changing ionization state of the
gas with distance from the Galactic plane.  Differences from element to
element in the 
destructive liberation of atoms from dust grains could also influence
the relative 
behavior of the z distributions of the highly ionized atoms.

Reviews of the many theories for the origins of the highly ionized atoms in
the ISM are found in Spitzer (1990,1996) and Sembach, Savage, \& Tripp (1997).
The three primary types of theories involve conductive heating (CH)
where cool gas
evaporates into an adjacent hot medium, radiative cooling (RC) where hot gas
cools as it flows into the
halo or down onto the disk, and turbulent mixing layers (TML) where hot and
cool gas are mixed through  turbulent entrainment (Slavin, Shull, \& Begelman
1993).  Table~2 of Spitzer (1996) provides a summary of the many
models along with the references to the theoretical literature.  The different
models make specific predictions for the expected values of
N(\ion{C}{4})/N(\ion{O}{6}).  The CH models are compatible with the
value of
N(\ion{C}{4})/N(\ion{O}{6}) $\sim$ 0.15
found at low $|z|$, while the larger values of $\sim$0.6 found toward
the extragalactic objects observed by FUSE are better explained by a
combination of the RC and TML models.

Shull \& Slavin (1994) developed a hybrid model for the highly ionized gas
in the Galactic halo in order to explain the smaller scale height of \ion{N}{5}
compared to \ion{Si}{4} and \ion{C}{4} suggested by the IUE and HST observations
available in 1994.   In their model the highly ionized ions at
low $|z|$ are produced mainly in isolated SNRs while those at high $|z|$
are mainly found in radiatively cooling superbubbles that break through the
disk producing Rayleigh-Taylor instabilities and turbulent mixing layers.
Possible support for the origin of the high ions at low $|z|$ in
isolated SNRs
follows from the detailed SNR modeling of Shelton (1998).   Confirmation
that the scale height difference (smaller scale heights for ions
with higher ionization potentials)  first seen for \ion{N}{5} and
\ion{C}{4} is also
clearly present in the new FUSE \ion{O}{6} measurements suggests that
such hybrid models
offer substantial promise for explaining the origin of the highly ionized
species in the Galactic halo.  Another example of a hybrid model is that
of Ito \& Ikeuchi (1988) which includes the cooling gas of a Galactic
fountain flow (Shapiro \& Field 1976) to provide the hot collisionally
ionized gas
and photoionization from the
extragalactic background (Hartquist, Pettini, \& Tallant 1984;
Fransson \& Chevalier 1985) to assist in the production of \ion{Si}{4}
and \ion{C}{4}.
The ionizing photons might also be provided by hot white dwarfs
(Dupree \& Raymond 1983).   The new observations with FUSE imply several
processes may be required to achieve a more complete understanding of the
origins of the low density highly ionized gas extending away from the
Galactic plane.

\acknowledgements
This work is based on data obtained for the Guaranteed Time Team by the
NASA-CNES-CSA FUSE mission operated by the Johns Hopkins University.
Financial support to U. S. participants has been provided by NASA contract
NAS5-32985.

\noindent
\begin{figure}
\includegraphics{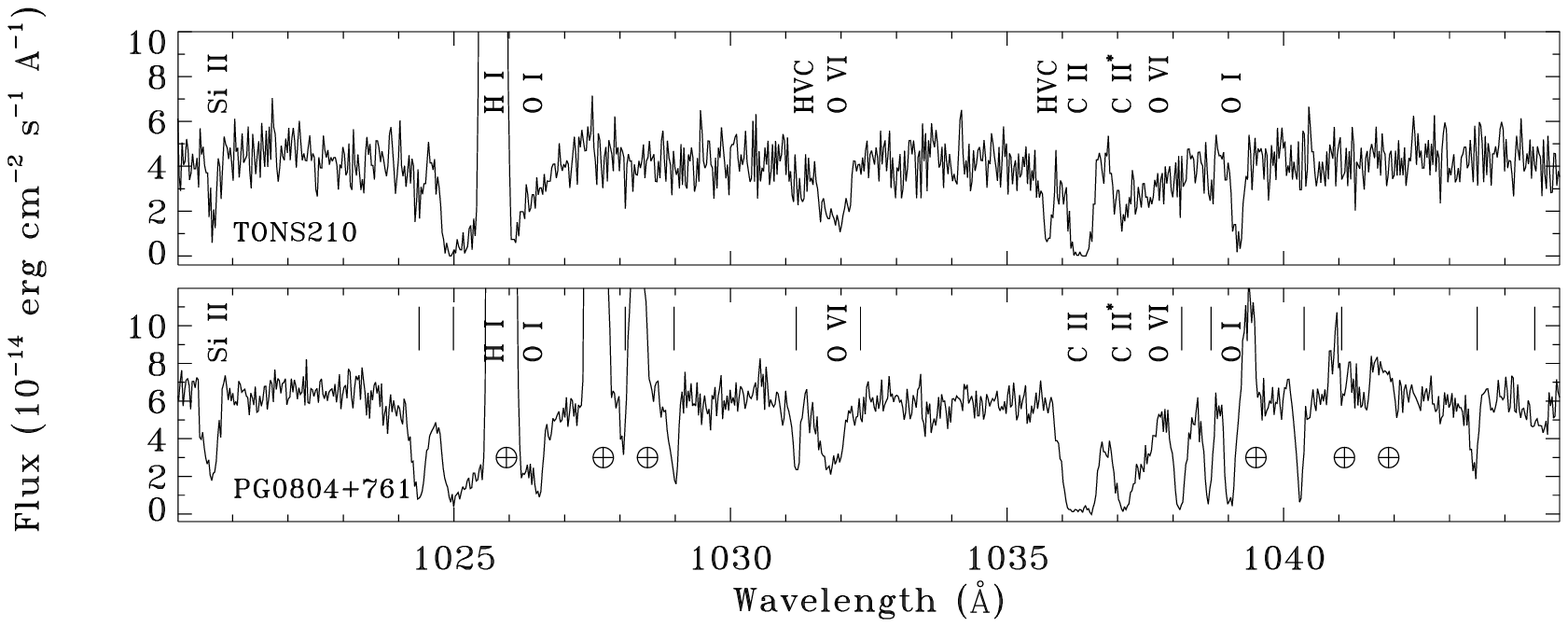}
\caption{Portions of the FUSE spectra of PG~0804+761 and Ton ~S210 over
the wavelength
range from 1020 --1045~\AA.  Galactic atomic absorption lines are
identified.  Galactic
H$_2$ lines are indicated with the tick marks.   The \ion{O}{6}
$\lambda1037.62$ line lies near strong absorption by \ion{C}{2}$^*$
$\lambda1037.02$ and the
H$_2$ (5--0) R(1) and P(1) lines at 1037.15 and 1038.16~\AA.  The
\ion{O}{6} 1031.93~\AA\ line
is usually relatively free of blending since the nearby H$_2$ (6--0)
R(3) and R(4) lines at
1031.19 and 1032.36~\AA\ are often  weak. The day and nighttime
integration for
PG~0804+761 shows airglow emission from \ion{O}{1} at the positions
marked with the $\oplus$.
Both integrations are contaminated by emission from terrestrial
Ly$\beta$ $\lambda1025.72$.}
\end{figure}

\begin{deluxetable}{lccccccccccc}
\tablecolumns{13}
\tablewidth{0pt}
\tablecaption{Results Based on the \ion{O}{6} $\lambda$1031.926 Line}
\tablehead{Object & V  & $z$ & $l$ & $b$ & T$_{exp}$ &
W$_\lambda\pm\sigma$ & v$_-$ & v$_+$ & log\,N$\pm\sigma$ &
log\,[Nsin$|b|$] & h\tablenotemark{a} \\
& & & ($^\circ$) & ($^\circ$) & (ksec) & (m\AA) &
\multicolumn{2}{c}{(\kms)} & (10$^{14}$ cm$^{-2}$) & & (kpc) }
\startdata
PG\,0052+251 & 15.4 & 0.155 & 123.9    & --37.4 & 16.8 & $<$200
(3$\sigma$) & --150 & +150 & $<$14.20\tablenotemark{b} & $<$13.98&$<$1.6\\
PG\,0804+761 & 15.2 & 0.100 & 138.3    & +31.0  & 39.6 & 286$\pm$19 &
--160 & \phn+80      & 14.50$\pm$0.04                  & 14.21&2.6\\
ESO\,141-55  & 13.6 & 0.037 & 338.2    & --26.7 & 35.8 & 256$\pm$21 &
--100 & +100         & 14.50$\pm$0.05                  & 14.15&2.3\\
VII\,Zw\,118 & 14.6 & \nodata   & 151.4    & +26.0  & 42.8 & 144$\pm$35
& --100 & +100         & 14.16$\pm$0.14              & 13.80&1.0\\
H\,1821+643  & 14.2 & 0.297 & \phn94.0 & +27.4  & 48.2 & 319$\pm$20 &
--175 & +100         & 14.57$\pm$0.05\tablenotemark{c} & 14.23&2.8\\
Mrk\,509     & 13.1 & 0.034 & \phn36.0 & --29.9 & 54.9 & 395$\pm$20 &
\phn--95 & +160      & 14.70$\pm$0.04\tablenotemark{c} & 14.40&4.1\\
Mrk\,876     & 15.5 & 0.129 & \phn98.3 & +40.4  & 45.9 & 223$\pm$15 &
--100 & \phn+60      & 14.38$\pm$0.04\tablenotemark{c} & 14.19&2.5\\
Ton\,S180    & 14.3 & 0.062 & 139.0    & --85.1 & 16.6 & 240$\pm$22 &
\phn--75 & +120      & 14.43$\pm$0.06\tablenotemark{c} & 14.43&4.3\\
Ton\,S210    & 15.2 & 0.117 & 225.0    & --83.2 & 13.5 & 369$\pm$36 &
--130 & +170         & 14.64$\pm$0.06\tablenotemark{c} & 14.64&7.0\\
PKS\,2155-304& 13.1 & 0.116 & \phn17.7 & --52.3 & 37.1 & 201$\pm$12 &
\phn--70 & +150      & 14.31$\pm$0.04\tablenotemark{c} & 14.21&2.6\\
NGC\,7469    & \nodata  & 0.016 & \phn83.1 & --45.5 & 29.9 & 117$\pm$18
& --100 & \phn+90  & 14.16$\pm$0.09\tablenotemark{c} & 14.01&1.7\\
\\
3C\,273      & 12.8 & 0.158 & 290.0    & +64.4 & \nodata & \nodata &
\nodata & \nodata & 14.84$\pm$0.11\tablenotemark{d} & 14.80&10.\\
\enddata
\tablenotetext{a}{The value of this scale height assumes
$n_0$(\ion{O}{6})$ = 2\times10^{-8}$ cm$^{-3}$. }
\tablenotetext{b}{Limit assumes a linear curve of growth. }
\tablenotetext{c}{The column density and equivalent widths do not
include contributions
from high velocity gas detected beyond the listed integration
range ($v_-$ to $v_+$).  This high velocity gas is discussed by Sembach
et al.\ (2000).}
\tablenotetext{d}{The value of N(\ion{O}{6}) for 3C\,273 is an ORFEUS
result from Hurwitz et al.\ (1998).}
\end{deluxetable}

\end{document}